\documentclass[epj,twocolumn]{webofc}
\usepackage[varg]{txfonts}   % Web of Conferences font
\usepackage{subfigure}
\woctitle{Powders \& Grains 2017}
\begin{document}
\title{Evolution of force networks in dense granular matter close to 
jamming
}
%
% subtitle is optional

\author{\firstname{Lou} \lastname{Kondic}\inst{1}\fnsep\thanks{\email{kondic@njit.edu }} \and
        \firstname{Miroslav} \lastname{Kram\'ar}\inst{2}\fnsep\thanks{\email{miroslav.kramar.1@gmail.com        
        }} \and
        \firstname{Lenka} \lastname{Koval\v cinov\'a}\inst{1}\fnsep\thanks{\email{lk58@njit.edu        
     }} \and
     \firstname{Konstanin} \lastname{Mischaikow}\inst{3}\fnsep\thanks{\email{mischaik@math.rutgers.edu}}     
}

\institute{Department of Mathematical Sciences,
New Jersey Institute of Technology,
University Heights,
Newark, NJ 07102, USA \and
           Advanced Institute for Materials Research, Tohoku University, Sendai, Japan\and
          Department of Mathematics and BioMaPS Institute,
Hill Center-Busch Campus,
Rutgers University,
110 Frelinghusen Rd,
Piscataway, NJ  08854, USA          }

\abstract{%
When dense granular systems are exposed to external forcing, they evolve on the time scale that is typically related to the externally imposed one (shear or compression rate, for example).   This evolution could be characterized by observing temporal evolution of contact networks. However, it is not immediately clear whether the force networks, defined on contact networks by considering force interactions between the particles, evolve on a similar time scale.  To analyze the
evolution of these networks, we carry out discrete element simulations of 
a system of soft frictional disks exposed to compression that leads to 
jamming.   By using the tools of computational topology, we show that 
close to jamming transition, the force networks evolve on the time scale 
which is much faster than the externally imposed one.  The presentation will 
discuss the factors that determine this fast time scale. }
\maketitle

\section{Introduction}

The interaction between particles that build dense granular matter (DGM) is 
characterized by the presence of a force network, mesoscopic structure
that spontaneously form as a granular system is exposed to shear
or compression.   These networks are crucial for understanding the 
static and dynamic properties of DGM, and have been 
recently studied extensively, using the tools 
that vary from statistical methods~\cite{peters05,tordesillas_pre10,tordesillas_bob_pre12}
to network type of analysis~\cite{daniels_pre12, herrera_pre11,walker_pre12} and
application of topological methods~\cite{arevalo_pre10, arevalo_pre13, epl12,pre13,physicaD14,pre14}.

Force networks exhibit complicated time dependent structures. 
Earlier approaches, used to analyze these structures, usually involved rather arbitrary choices of the force thresholds and often depended on ad hoc descriptions of the structures. 
To avoid these problems we define the force network \cite{physicaD14} as a collection of sets $\{ X_\theta\}_{\theta \in \mathbb R}$ such that for $\theta \geq 0$ the set  $X_\theta$ is the part of the contact network on which the force interactions between the particles exceed the value $\theta$. 
In the series of papers~\cite{epl12,pre13,physicaD14,pre14},  we make use of an algebraic topological technique that rigorously describes geometric structures of the sets $\{ X_\theta\}_{\theta \in \mathbb R}$ over all force levels $\theta$. This approach provides a  quantitative description of the force networks. Moreover, it allows us to define a concept of distance between force networks, so we can quantitatively compare them.

In~\cite{pre14} we focused on comparing force networks in simulated granular systems built from soft, possibly inelastic and/or frictional particles  exposed to steady slow compression.  In that work we found that force networks' evolution was characterized by large distances between consecutively sampled states of the systems that were approaching jamming transition from below, and the evolution was much smoother for 
jammed systems.   In particular close to jamming transition, we found rather dramatic
evolution of the force networks in all considered systems (frictional or not, with 
elastic or inelastic particles).  

The following unexpected findings obtained in~\cite{pre14} motivate the present peper.
Distances between force networks at consecutive states of the system sampled at the rate used in~\cite{pre14} are comparable to the distances between force networks at the same packing fraction obtained from different realizations, see Figure 1. Sec.~\ref{sec:methods} provides a brief description of the plotted quantities while more in depth discussion of the figure is presented in Sec.~\ref{sec:results}. This result suggests that force networks evolve on a time scale that is much faster than the inverse sampling rate and therefore, the main topic of the present paper is the analysis of the system on extremely fast time scales.  Furthermore, we  discuss how the time scale characterizing  evolution of force networks depend on two externally imposed  time scales: the slow one imposed by  external driving (compression in the case considered here) and the fast one imposed by the temporal discretization used in our simulations.  We show that variation of these two scales provides further insight into the mechanisms governing force network evolution.  

\section{Methods}
\label{sec:methods}
Here we give a brief overview of the techniques used; the reader is referred to~\cite{pre14} for 
more detailed description of the protocol and definitions of various quantities.    We consider
a square domain in two spatial dimensions exposed to slow compression by the moving walls. 
The particles are polydisperse, soft, inelastic disks interacting by normal and tangential forces, with the latter
ones modeled by Cundall-Strack type of interaction~\cite{cundall79}.  
All quantities
are expressed using the average particle diameter, {\bf $d$}, as
the length scale, the binary particle collision time $\tau_c = \pi
\sqrt{d/(2 g k_n)}$ as the time scale, and the average particle mass,
$m$, as the mass scale.  Here  $k_n$ (in units
of ${ m g/d}$) is the spring constant used for modeling (linear) repulsion force between colliding particles,
and $g$ is gravity (used only for the purpose of scaling).  
For the initial configuration, particles ($\approx 2000$)  are placed on a square
lattice and given random initial velocities.  Except if specified differently,   the walls move at a uniform
inward velocity $v_{c}= 2.5\cdot 10^{-5}$.  We integrate
Newton's equations of motion for both the translational and rotational
degrees of freedom using a $4$th order predictor-corrector method with
time step $\delta t = {1/50}$. 

We use persistent homology \cite{edelsbrunner:harer} to quantitatively describe the structure of force networks. 
Briefly, each network is represented by two persistence diagrams that quantify the changes in geometry of the sets $X_\theta$ for all force levels $\theta$. 
The $\beta_0$ ($\beta_1$) persistence diagram encodes the appearance and disappearance of connected components (loops) by recording all values at which a connected component (loop) appears in a set $X_\theta$ and all values of the force level at which the corresponding connected component (loop) disappears.
For a rather brief discussion of persistence, we refer the reader to~\cite{pre14} and for a more in-depth presentation to~\cite{physicaD14}. 
There are different types of well defined distances between the persistence diagrams. Each of these distances measures differences between the force levels at which distinct topological features appear and disappear. In particular, the distance between two $\beta_0$ persistence diagrams corresponding to different force networks measures differences between the force values at which distinct connected components/clusters are formed and merged together. In a loose sense, this corresponds to measuring differences between the force chains.
The results reported in this paper do not depend on the particular choice of the distance and thus, for brevity, we will focus only on the so called Wasserstain distance, $d_{W^1}$, between the $\beta_0$ persistence diagrams (see~\cite{physicaD14} for a detailed definition).

\section{Results}\label{sec:results}

We start by recalling the results obtained for slow sampling rate of approximately $100$ samples covering the packing fraction range  $[0.63, 0.9]$.  Since the evolution of
force networks varies from realization to realization, and the distances between 
considered states of the system may be rather noisy, we considered the averages over a large number ($20$) of realizations~\cite{pre14}.  Figure~1 shows that the  distances between two consecutive states of a  given realization  are  comparable to the distances between two  states  resulting from different realizations. This  suggests that the memory of the system is lost on the time scale faster than  the one defined  by the slow sampling rate.

\begin{figure}
\centering
{\includegraphics[width=2.5in]{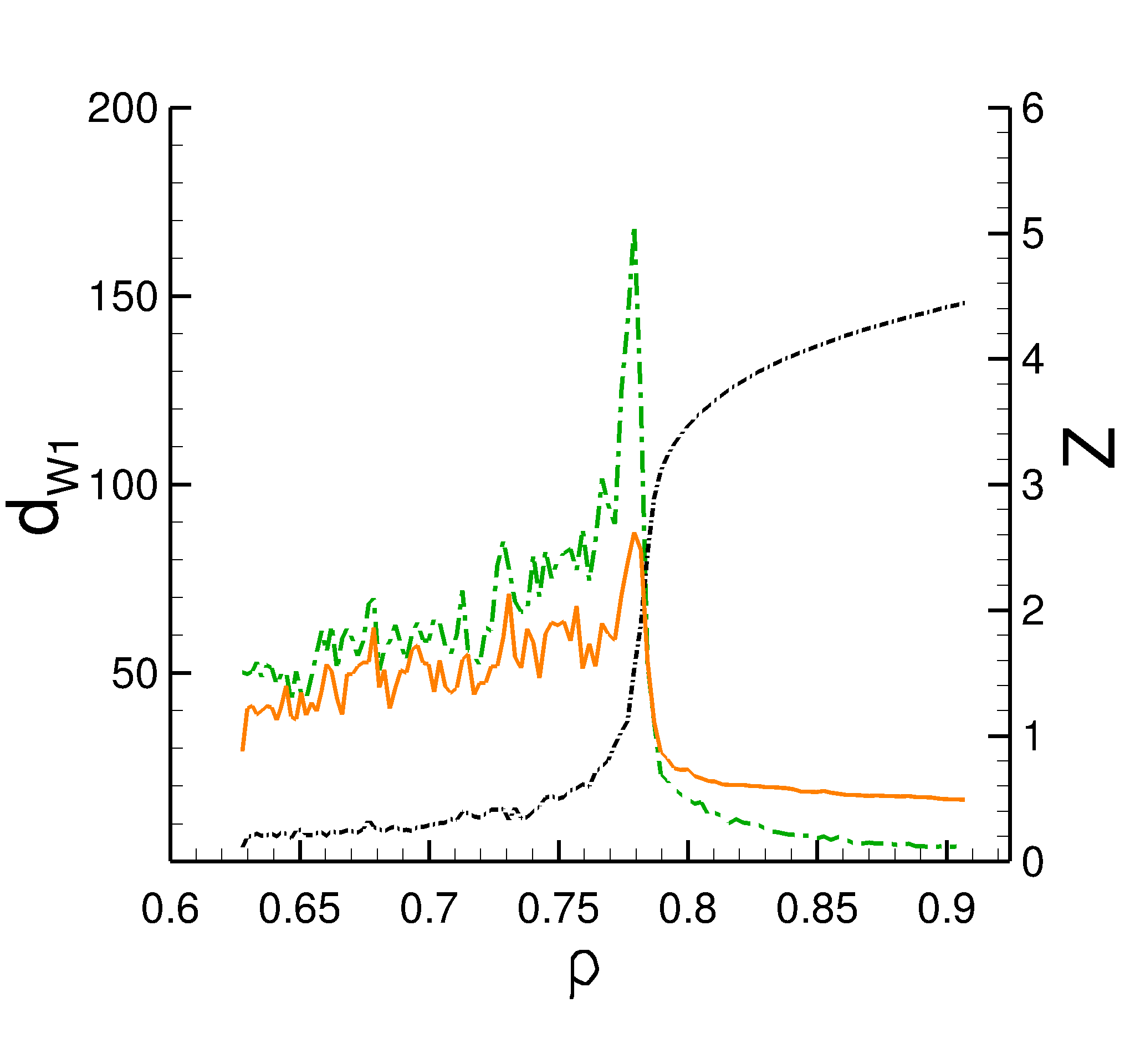}
\caption{Wasserstein distance, $d_{W^1}$, measuring differences between force networks.  The dash-dotted (green) line shows the distances of the networks 
averaged over $20$ realization and the solid (orange) line shows the distances between different realizations.  The dotted (black) line  shows the number of contacts, $Z$; the rapid rise is a sign of jamming transition.
}}
\label{fig:slow}
\end{figure}

\begin{figure}[thb]
\centering
\includegraphics[width=3in]{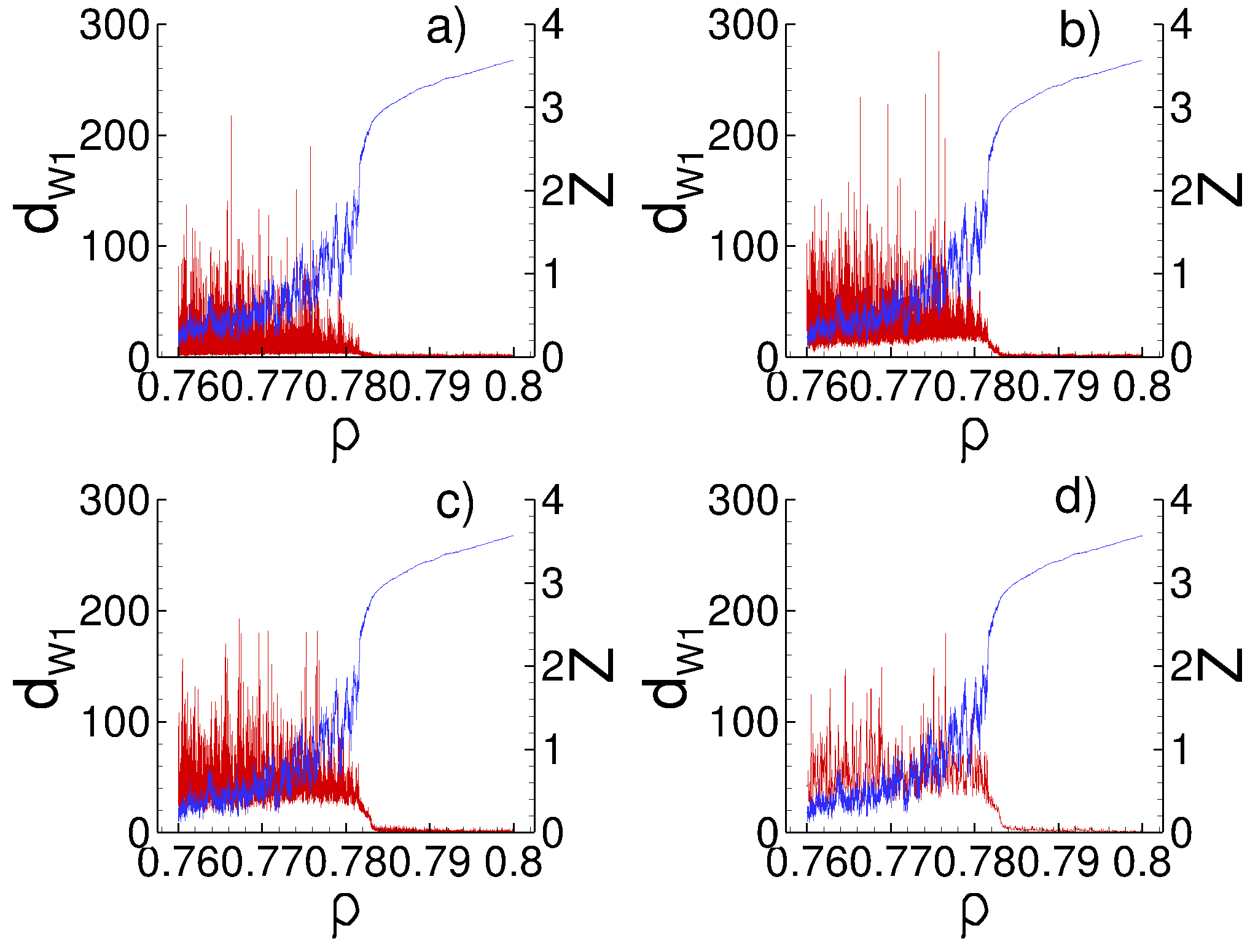}
 \caption{Wasserstein distance $d_{W^1}$ (red) between  $\beta_0$ persistence diagrams computed for different sampling rates: a) $r_1$ b) $r_{10}$, c) $r_{100}$ and d) 
$r_{1000}$.   The number of contacts between particles, $Z$ is shown in blue. }\label{fig:fast_W_p1}
\end{figure}

To proceed, we consider much faster sampling rates.
We  also focus on presenting and analyzing the results from a single realization, since averaging  the results  over realizations obscures  part of the information about force network evolution in a given system.   
Figure~\ref{fig:fast_W_p1}  shows $d_{W^1}$ distance
for four sampling rates $r_1, r_{10}, r_{100}$ and $r_{1000}$;
here the index of a rate corresponds to 
the number of computational steps of duration $\delta t$ between two consecutive sample points.
The difference  in the packing fraction, $\rho$, between two consecutive states sampled at  the 
rate $r_i$ is $\Delta \rho = i \cdot \delta\rho$ where $\delta\rho  \approx 4\cdot 10^{-8}$ is the change of $\rho$ 
during one computational step.   Note that $r_1$ corresponds to the fastest possible sampling rate, at which the state
of the system is recorded at every time step.  First, we note that, as expected, for faster sampling rates, there is a large number
of points such that the distance is very small, suggesting that the force networks do not evolve much during
a significant portion of the data shown.   However, contrary to expectations,  there is  an {\it increase} of the 
peaks' sizes as the sampling rate is increased from  $r_{1000}$ to $r_{100}$ and $r_{10}$, 
followed by a gentle decrease as the rate is increased from $r_{10}$ to $r_1$. 
This essentially means that one needs to sample at the rate comparable to $r_{10}$ (at which the peaks in $d_{W^1}$ 
distance are the largest) to capture the corresponding large events  present in networks' evolution.  In turn, this implies
that  large changes between the force networks occur on the time scales shorter than $\tau_c$
(capturing the system evolution with the sampling rate given by $1/\tau_c$ would correspond to $r_{50}$).

\begin{figure}[thb]
\centering
{\includegraphics[width = 2.7in]{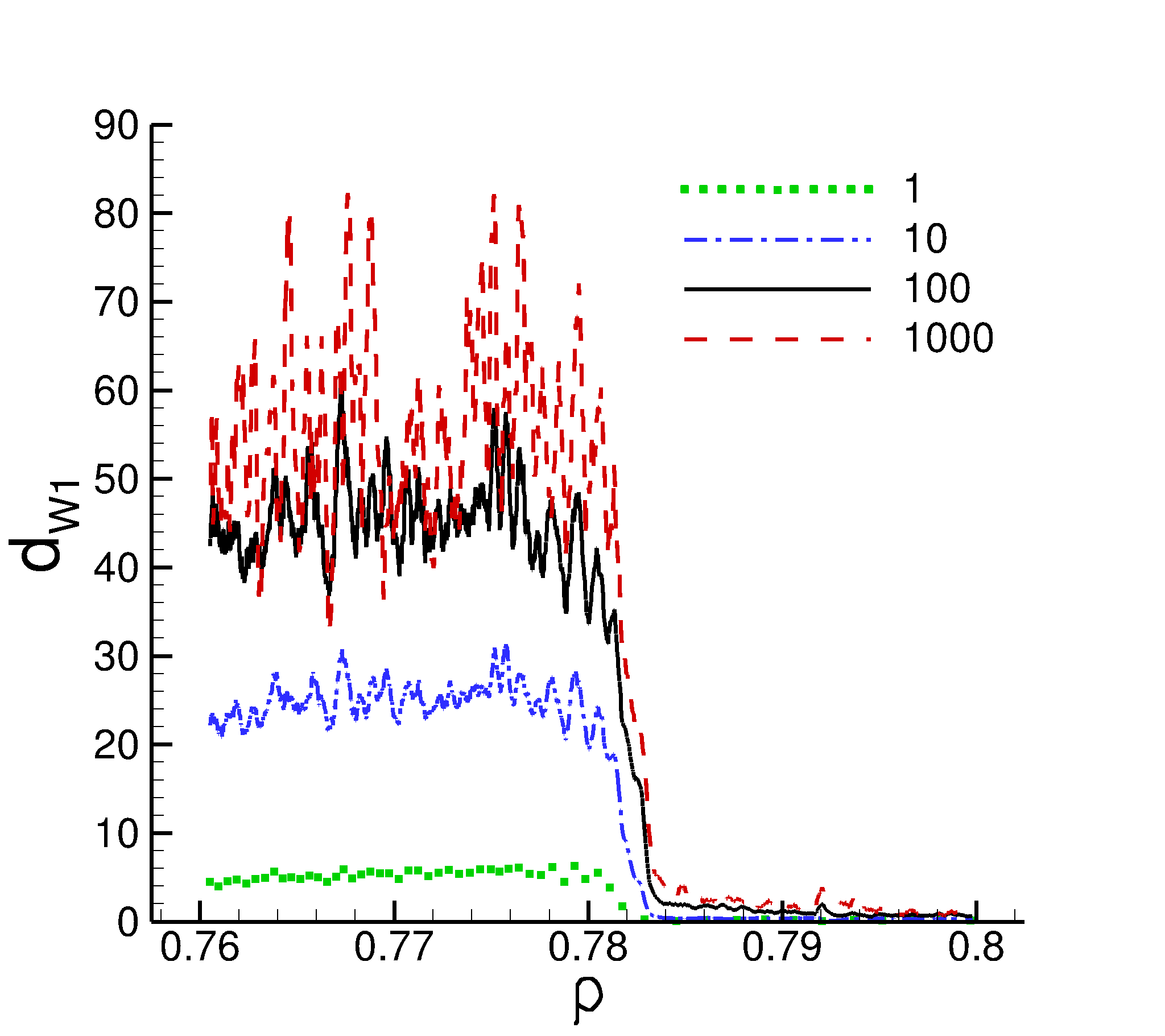}}
\caption{ Running average of $d_{W^1}$-distances over  $10^4 \delta \rho$. The curves correspond to the sampling rate used to compute the distances 
shown in  Fig.~\ref{fig:fast_W_p1}.}
\label{fig:dist_ave_B0}
\end{figure}

The number of data points shown in Fig.~\ref{fig:fast_W_p1} is large;  e.g., Fig.~\ref{fig:fast_W_p1}a shows about $10^6$ data points.  
Since it is impractical to work with such a large number of data points, we proceed by carrying out a running average of the data 
shown in Fig.~\ref{fig:fast_W_p1}, and then proceed with considering a subset of the data extracted from the region 
very close to the jamming transition (that occurs
at $\rho \approx 0.78$ for the present system).  
Figure~\ref{fig:dist_ave_B0} shows the running averages for $d_{W^1}$ distance. The averaging interval is given by
$10^4 \delta \rho \approx 4\cdot 10^{-4}$. 
This averaging interval corresponds to the averaging time of about $200\tau_c$ which for a 
common granular  system may range between a $\mu$sec and  a msec. The actual number of points used for computing 
varies from $10^4$ for the fastest sampling rate $r_1$ to $10$ for the slowest sampling rate $r_{1000}$. 

Before jamming transition a typical distance between consecutive states grows as the sampling rate decreases, but at a rate that is typically
much slower than expected.   Note that the consecutive 
sampling rates differ by one order of magnitude. However, the differences between  averaged distances is one order of magnitude 
only for the sampling rates $r_1$ and $r_{10}$ and becomes  significantly smaller for slower sampling;  the results are 
actually overlapping for
 $r_{100}$ and $r_{1000}$.   This  finding suggests, consistently with the results shown in Fig.~\ref{fig:fast_W_p1},  that the time scale 
 on which force networks evolve is comparable to $\tau_c$ or even shorter.

The fast time scale on which force networks evolve raises new questions.
Recall that $\delta t$ is the time interval needed for information about possible interactions to travel across a particle and hence $\delta t$ is related to the speed of sound, $v_s$, in the material.
In particular, the approximate speed of sound, determined by the numerical time step, is given (in physical units) by $d/({\delta t \tau_c})$, and therefore, information about the features present  in force networks that evolve on the time scale given by $\delta t$ essentially propagates by the speed of sound.
Thus, it is natural to ask whether the results may be influenced by $\delta t$.  

\begin{figure}[thb]
\center
\includegraphics[width=3in]{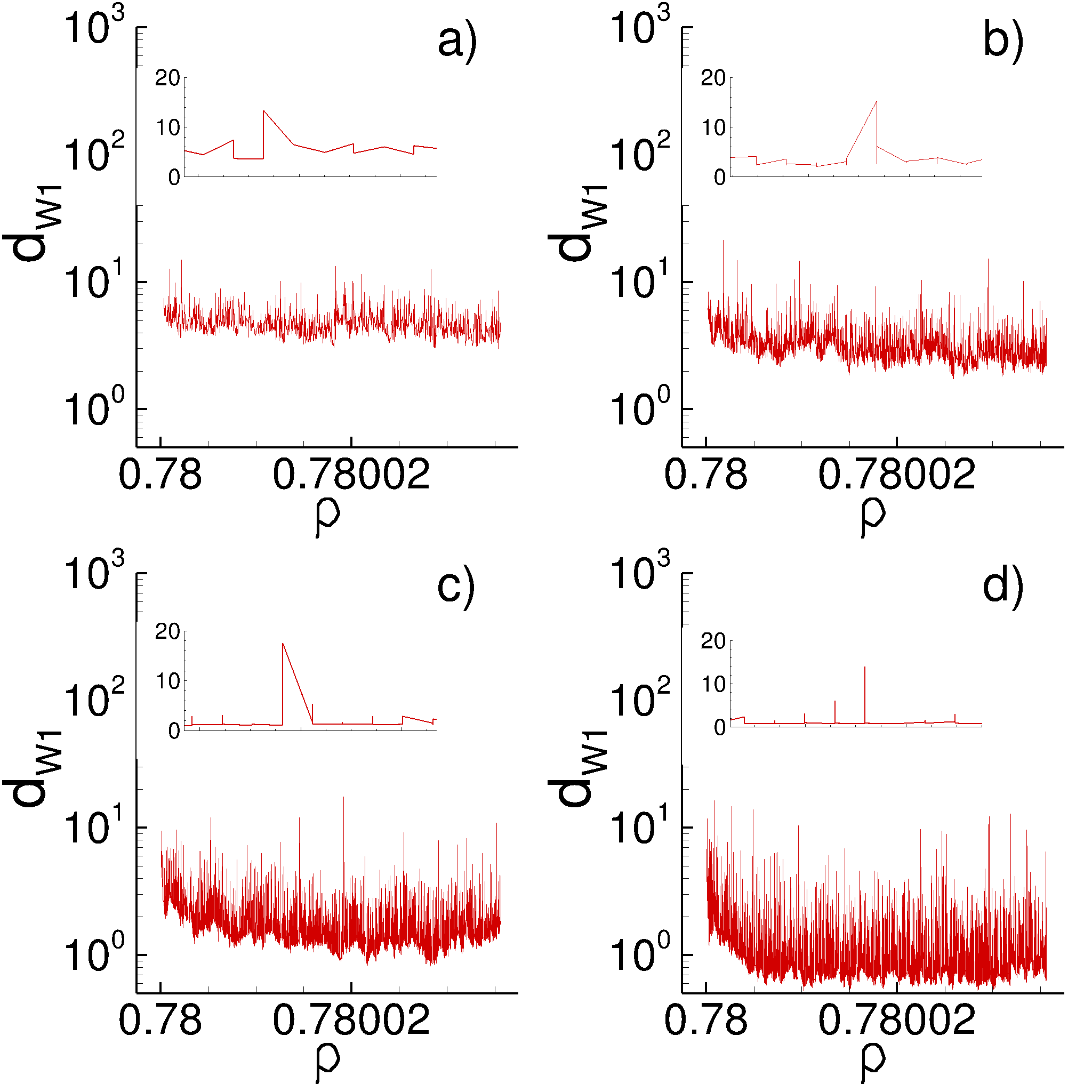}
\caption{
$d_{W^1}$ distance between $\beta_0$ persistence diagrams corresponding to the consecutive force networks (one computational time step apart) for $\rho \approx 0.78$ as
$\delta t$ is reduced, 
with the part a) using the reference value $\delta t = 1/50$, and the parts b), c), d) using $\delta t/2$, $\delta t/4$, and $\delta t/8$, 
respectively.   
The insets show the distances for  few consecutive time steps, with the shown range of $5\times 10^{-7}$.
The compression speed, $v_c$, is kept fixed.
}
\label{fig:dt}
\end{figure}

Figure~\ref{fig:dt} shows an influence of decreasing  $\delta t$  on $d_{W^1}$ distance  in a small interval based at $\rho \approx 0.78$. Consequences of decreasing the $\delta t$ are two fold.
On one hand, 
as $\delta t$ decreases, on average  $d_{W^1}$ decreases (the number of points at which 
$d_{W^1}$ is very small increases as $\delta t$ decreases). On the other hand, we notice that  the peaks' heights do not necessarily
decrease with $\delta t$, and more importantly, the peaks still relax on the scale corresponding 
to $\delta t$: for smaller $\delta t$'s, the peaks relax faster (see the insets in Fig.~\ref{fig:dt}).
This finding is consistent with the general discussion above regarding the relevance of the speed of
sound, and is showing clearly that in the considered system the force networks evolve on the time scale 
determined by the speed of sound in the material.  

Another question to ask is related to the influence of the compression speed, $v_c$, which defines a slow time 
scale in the problem.   As $v_c$ is reduced, one may expect that the distance between the consecutive states of the system
becomes smaller.   As we show next, such trend is indeed observed in our results.

\begin{figure}[thb]
\center
\includegraphics[width=3in]{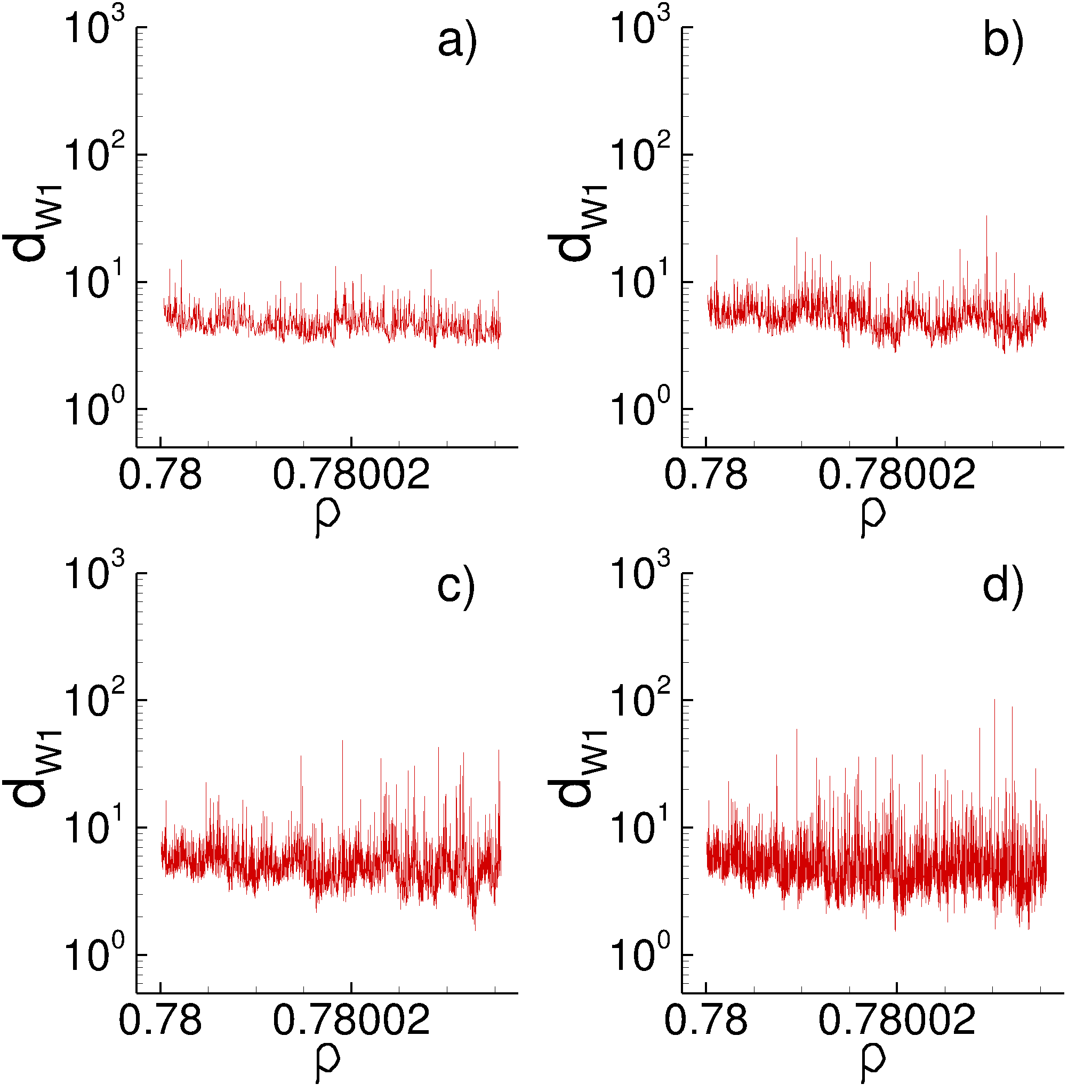}
\caption{
$d_{W^1}$ distance between $\beta_0$ persistence diagrams corresponding to the consecutive force networks (one $\delta t$ apart) for $\rho \approx 0.78$ 
as the compression speed is reduced, with the part a) 
using the reference value, $v_c$, and the parts b), c), and d) using $v_c/2$, $v_c/4$ and $v_c/8$, respectively.
$\delta t$ is kept fixed. 
}
\label{fig:vc}
\end{figure}

Figure~\ref{fig:vc} shows $d_{W^1}$ distance at the same $\rho$ as in Fig.~\ref{fig:dt}, but now for four (decreasing)
$v_c$'s.    We see that, indeed, on average $d_{W^1}$ decreases as $v_c$ is decreased, meaning that there are many
$\delta t$ intervals when the force networks essentially do not evolve.     However, 
the height of the peaks does not necessarily decrease with $v_c$; we even find examples of larger peaks for slower 
compression.   This finding can be 
understood by realizing that as the system is compressed slower, the particles have  higher chance to rearrange,
leading in some cases to large changes in the force network and increased distances between the consecutive states.

\section{Conclusions} 
\label{sec:conclusions}

The force networks are found to evolve on the time scale  that is much 
faster than the one introduced by externally imposed driving.  As a consequence, an
analysis of dynamics of force networks based on slow sampling rates
may provide incomplete information about the networks' evolution, since slow sampling may miss large events
that occur on faster time scales.   In particular, we find that the time scale on which evolution occurs is 
specified by $d/v_s$, where $d$ is a typical particle diameter and $v_s$ the speed of sound in the material.   It will 
be of interest to see to which degree such behavior can be seen in experimental systems.   Furthermore, it 
should be noted that in the present work we focus on the systems very close to jamming transition; it remains
to be seen whether such fast evolution as uncovered here may be seen further away from jamming, and/or for
different type of external forcing.   

The reported results rely on the use of computational homology that allows us to quantify the differences 
between force networks as a granular system evolves. One particular strength of the computational homology 
approach is that it is system-independent and can be applied to any system, independently of the physical properties of the considered system, such as
the type of interaction between the particles, or the number of dimensions of the physical space in which they live.   Therefore, natural
next step will be to consider particulate systems characterized by different physical properties (friction, polydispersity, 
cohesion, shape) or different geometry (2D versus 3D).  Furthermore, the approach can be as well applied to experimental
systems, such as those built from photoelastic particles, where detailed information about the force networks is
available.    Further analyses of such systems will be the subject of our future works.  
\vspace{0.1in}

\section*{Acknowledgements}
This work was supported by the NSF Grant DMS-1521717 and by DARPA contract No. HR0011-16-2-0033.

%\vspace{0.1in}
%\bibliography{granulates}

\end{document}